\documentstyle[sprocl]{article}

\input{psfig}

\def\be{\begin{equation}}
\def\ee{\end{equation}}
\def\bea{\begin{eqnarray}}
\def\eea{\end{eqnarray}}

\begin{document}

\title{COLLECTIVE STATES FROM RANDOM INTERACTIONS}

\author{R. BIJKER$^1$ and A. FRANK$^{1,2}$}

\address{$^1$ ICN-UNAM, A.P. 70-543, 04510 M\'exico, D.F., M\'exico\\
$^2$ CCF-UNAM, A.P. 139-B, Cuernavaca, Morelos, M\'exico} 

\maketitle

\abstracts{The anharmonic vibrator and rotor regions in nuclei are 
investigated in the framework of the interacting boson model using 
an ensemble of random one- and two-body interactions. Despite the 
randomness of the interactions (in sign and size) we find a 
predominance of $L^P=0^+$ ground states and strong evidence for the 
occurrence of both vibrational and rotational band structure.} 

\section{Introduction} 

In empirical studies of medium and heavy even-even nuclei very 
regular features have been observed, such as the 
tripartite classification of nuclear structure into seniority, 
anharmonic vibrator and rotor regions \cite{Zamfir}. 
In each of these three regimes, the energy systematics is extremely 
robust, and the transitions between different regions occur very 
rapidly, typically with the addition or removal of only one or two 
pairs of nucleons. These robust features suggest that there exists 
an underlying simplicity of low-energy nuclear structure never 
before appreciated. In order to address this point we present the 
results of a study of the systematics of collective levels in the 
framework of the interacting boson model (IBM) with random 
interactions \cite{BF}. 

\section{Randomly interacting bosons}

In the IBM, collective excitations in nuclei are described in 
terms of a system of $N$ interacting monopole and quadrupole 
bosons \cite{IBM}. We consider the most general one- and two-body 
Hamiltonian $H= H_1/N + H_2/N(N-1)$. The two one-body and seven 
two-body matrix elements of the IBM are 
chosen independently from a Gaussian distribution of random 
numbers with zero mean and variance $v^2$, such that the 
ensemble is invariant under orthogonal basis transformations. 
In order to remove the $N$ dependence of the many-body matrix 
elements, we have scaled $H_1$ by $N$ and $H_2$ by $N(N-1)$. 
In all calculations we take $N=16$ bosons and 1000 runs. 
For each set of randomly generated one- and two-body matrix 
elements we calculate the entire energy spectrum and the $B(E2)$ 
values between the yrast states. 

Just as for the nuclear shell model with random interactions 
\cite{JBD,random}, we find a predominance of $L^P=0^+$ ground 
states. In 63.4 $\%$ of the runs the ground state has $L^P=0^+$, 
followed by the state with the maximum value of the angular 
momentum $L^P=32^+$ with 16.7 $\%$ and $L^P=2^+$ with 13.8 $\%$. 

\begin{figure}
\centerline{\hbox{\psfig{figure=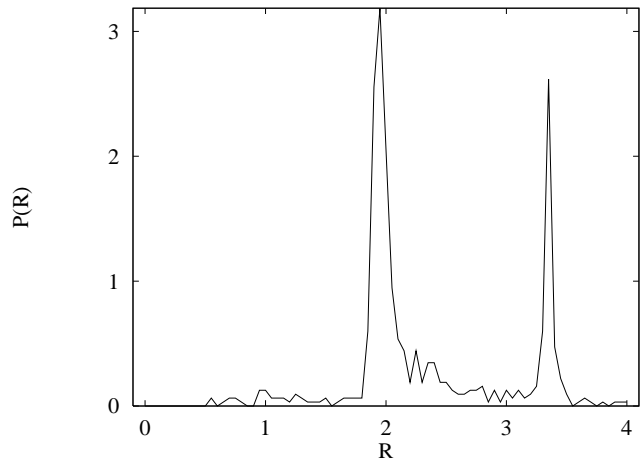,width=0.75\linewidth} }}
\caption[]{Probability distribution $P(R)$ of the energy ratio 
$R=[E(4^+)-E(0^+)]/[E(2^+)-E(0^+)]$ with $\int P(R) dR = 1$ 
in the IBM with random one- and two-body interactions. The 
number of bosons is $N=16$.}
\label{ratio12}
\end{figure}

For the cases with a $0^+$ ground state we calculate the 
probability distribution $P(R)$ of the energy ratio of the yrast 
states $R = [E(4^+)-E(0^+)]/[E(2^+)-E(0^+)]$. 
This energy ratio has characteristic values of $R\approx1$, $2$ 
and $10/3$ for the seniority, vibrational and rotational regions, 
respectively. Fig.~\ref{ratio12} shows a remarkable result: 
$P(R)$ has two very pronounced peaks, 
one at $R \sim 1.95$ and a narrower one at $R \sim 3.35$ \cite{BF}. 
These values correspond almost exactly to the harmonic vibrator 
and rotor values of 2 and 10/3 (see Table~\ref{BE2}). 

\begin{table}
\centering
\caption[]{Energies and $B(E2)$ values in the dynamical symmetry 
limits of the IBM \protect\cite{IBM}. In the $U(5)$ and $SO(6)$ 
limits we show the result for the leading order contribution to the 
rotational spectra.}
\label{BE2}
\vspace{15pt}
\begin{tabular}{|ccc|}
\hline
& & \\
& $R=\frac{E(4^+)-E(0^+)}{E(2^+)-E(0^+)}$  
& $\frac{B(E2;4^+ \rightarrow 2^+)}{B(E2;2^+ \rightarrow 0^+)}$ \\
& & \\
\hline
& & \\
$U(5)$  & $2$ & $\frac{2(N-1)}{N}$ \\
$SO(6)$ & $\frac{5}{2}$ 
& $\frac{10(N-1)(N+5)}{7N(N+4)}$ \\
$SU(3)$ & $\frac{10}{3}$ 
& $\frac{10(N-1)(2N+5)}{7N(2N+3)}$ \\
& & \\
\hline
\end{tabular}
\end{table}

\begin{figure}
\centerline{\hbox{\psfig{figure=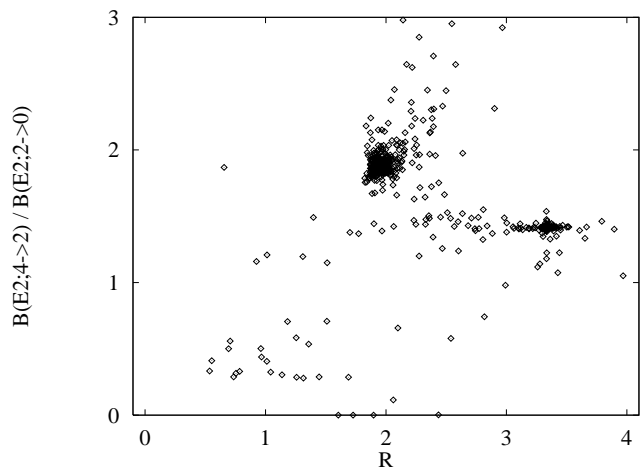,width=0.75\linewidth} }}
\caption[]{Correlation between ratios of $B(E2)$ values and energies 
in the IBM with random one- and two-body interactions.}
\label{corr}
\end{figure}

Energies by themselves are not sufficient to decide whether 
or not there exists a collective structure. Levels belonging to a 
collective band are connected by strong electromagnetic 
transitions. In Table~\ref{BE2} we show the energy ratio $R$ 
and the ratio of $B(E2)$ values of the $4^+ \rightarrow 2^+$ and 
$2^+ \rightarrow 0^+$ transitions for the three symmetry limits 
of the IBM \cite{IBM}. In the large $N$ limit, the ratio of 
$B(E2)$ values is 2 for the harmonic vibrator and 10/7 both for 
the $\gamma$-unstable rotor and the axially symmetric rotor. 
In Fig.~\ref{corr} we show a correlation plot between the ratio 
of $B(E2)$ values and the energy ratio $R$. There is a strong 
correlation between the first peak in the energy ratio and the 
vibrator value for the ratio of $B(E2)$ values (the concentration 
of points in this region corresponds to about 50 $\%$ of all 
cases), and for the second peak and the rotor value (about 25 
$\%$ of all cases) \cite{BF}. 

A subsequent study \cite{BF2} of the dependence of these collective 
features on the number of bosons $N$ and the rank $k$ of the 
interactions has shown that the vibrational and rotational band 
structures appear gradually as a function of $N/k$. For $N \sim k$ 
there is little or no evidence for such bands. For larger 
values of $N$ we first see the development of vibrational 
structure, followed later by the appearance of rotational bands.  
If $N$ increases further these band structures become more and more 
pronounced. 

\section{Conclusions}

A study of the IBM with random ensembles 
of one- and two-body Hamiltonians has shown that despite the 
randomness of the interactions the ground state has $L^P=0^+$ 
in 63.4 $\%$ of the cases. For this subset, the analysis of 
energies and quadrupole transitions shows strong evidence for the 
occurrence of both vibrational and rotational band structure. 
The inclusion of random three-body interactions does not 
significantly change the basic properties.

These regular features arise from a much wider class of Hamiltonians 
than are usually considered to be `realistic', and 
represent general and robust properties of the many-body dynamics, 
which enters via the reduction formulas for the $N$-body matrix 
elements of $k$-body interactions. Since the structure of the 
model space is completely determined by the corresponding degrees 
of freedom, these results emphasize the importance of the selection 
of the relevant degrees of freedom. 

A similar situation has been observed in the context of the nuclear 
shell model with respect to the pairing properties \cite{JBDT}. 

\section*{Acknowledgements}

This work was supported in part by DGAPA-UNAM under project IN101997, 
and by CONACyT under projects 32416-E and 32397-E.


\section*{References}

\begin{thebibliography}{99}

\bibitem{Zamfir}
N.V. Zamfir, R.F. Casten and D.S. Brenner, 
{\em Phys. Rev. Lett.} {\bf 72}, 3480 (1994).

\bibitem{BF}
R. Bijker and A. Frank, 
{\em Phys. Rev. Lett.} {\bf 84}, 420 (2000).

\bibitem{IBM}
F. Iachello and A. Arima, 
{\it The interacting boson model}  
(Cambridge University Press, 1987).

\bibitem{JBD} 
C.W. Johnson, G.F. Bertsch and D.J. Dean, 
{\em Phys. Rev. Lett.} {\bf 80}, 2749 (1998).

\bibitem{random}
R. Bijker, A. Frank and S. Pittel, 
{\em Phys. Rev.} C {\bf 60}, 021302 (1999).

\bibitem{BF2}
R. Bijker and A. Frank, 
{\em Phys. Rev.} C, in press. Preprint nucl-th/0004002. 

\bibitem{JBDT} 
C.W. Johnson, G.F. Bertsch, D.J. Dean and I. Talmi, 
{\em Phys. Rev.} C {\bf 61}, 014311 (2000). 

\end{thebibliography}
\end{document}